\documentclass[twocolumn, times, twocolappendix]{aastex701}
\usepackage{newtx}
\newcommand{\msun}{M$_\odot$}

\newcommand{\kms}{km\,s$^{-1}$}

\DeclareRobustCommand{\VAN}[3]{#2}
\let\VANthebibliography\thebibliography
\def\thebibliography{\DeclareRobustCommand{\VAN}[3]{##3}\VANthebibliography}

\graphicspath{{./}{figures/}}
\begin{document}

\title{IK~Pegasi and the Double-merger Path to Type Ia Supernovae}

\correspondingauthor{Na'ama Hallakoun}
\email{naama.hallakoun@weizmann.ac.il}
\correspondingauthor{Sahar Shahaf}
\email{sashahaf@mpia.de}

\author[orcid=0000-0002-0430-7793,sname=Hallakoun]{Na'ama Hallakoun}
\altaffiliation{These authors contributed equally.}
\affiliation{Department of Particle Physics and Astrophysics, Weizmann Institute of Science, Rehovot 7610001, Israel}
\email{naama.hallakoun@weizmann.ac.il}

\author[orcid=0000-0001-9298-8068,sname=Shahaf]{Sahar Shahaf}
\altaffiliation{These authors contributed equally.}
\affiliation{Max-Planck-Institut f\"ur Astronomie (MPIA), K\"onigstuhl 17, 69117 Heidelberg, Germany}
\email{sashahaf@mpia.de}

\author[0000-0001-6760-3074]{Sagi Ben-Ami}
\affiliation{Department of Particle Physics and Astrophysics, Weizmann Institute of Science, Rehovot 7610001, Israel}
\email{sagi.ben-ami@weizmann.ac.il}

\author[0009-0002-9137-0631]{Oren Ironi}
\affiliation{Department of Particle Physics and Astrophysics, Weizmann Institute of Science, Rehovot 7610001, Israel}
\email{oren.ironi@weizmann.ac.il}

\author[0009-0001-3501-7852]{Param Rekhi}
\affiliation{Department of Particle Physics and Astrophysics, Weizmann Institute of Science, Rehovot 7610001, Israel}
\email{param.rekhi@weizmann.ac.il}

\author[0000-0003-4996-9069]{Hans-Walter Rix}
\affiliation{Max-Planck-Institut f\"ur Astronomie (MPIA), K\"onigstuhl 17, 69117 Heidelberg, Germany}
\email{rix@mpia.de}

\begin{abstract}
Recent Gaia astrometry has revealed thousands of main-sequence $+$ white-dwarf binaries (MS$+$WD) at separations of ${\sim}\,0.1$--$10$\,au, including a subset hosting unusually massive ($\gtrsim 0.8$\,\msun) WDs. We argue that \textit{s}-process enrichment in the non-degenerate companion provides a powerful diagnostic for identifying WDs that formed via mergers in hierarchical triple systems. For a massive WD, standard single-star evolution requires a massive ($\gtrsim 4$\,\msun) progenitor, yet such progenitors produce negligible \textit{s}-process yields. We define \emph{IK~Peg--type} systems as those exhibiting this mass--yield tension: barium-enhanced companions orbiting WDs too massive to have descended from efficient \textit{s}-process producers. The well-known system IK~Peg exemplifies this class. Applying this framework to published spectroscopic data reveals several additional candidates, and we estimate that a few dozen such systems should exist in the current Gaia sample. If these systems trace inner-binary mergers in primordial triples, they represent observable intermediate stages toward eventual Type~Ia supernovae via the double-merger pathway, as predicted by recent population-synthesis models.
\end{abstract}


\section{Introduction}
\label{sec:intro}

Under standard single-star evolution, massive white dwarfs (WDs; $\gtrsim 0.8$\,\msun) descend from progenitors with masses of $4-8$\,\msun\ \citep[e.g.,][]{Cunningham_2024}. Observations indicate that such massive remnants comprise roughly one-third of single WDs in the local field population \citep{Hallakoun_2024}. In contrast, assuming a standard \citeauthor{Kroupa_2001} initial mass function \citep[IMF;][]{Kroupa_2001}, their progenitors are predicted to constitute only ${\lesssim}\,10\%$ of a 10\,Gyr-old WD progenitor population. This observed excess reinforces growing evidence that a substantial fraction of massive WDs are the products of binary mergers rather than isolated stellar evolution \citep[see][and references therein]{Hallakoun_2024}.

Recent Gaia astrometry has revealed a few thousand detached main-sequence $+$ white-dwarf (MS$+$WD) binaries with orbital separations of order ${\sim}\,1$\,au and a broad distribution of eccentricities \citep{Shahaf_2023, Shahaf_2024}. Notably, many of these systems exhibit no significant trend toward circularization, a characteristic typically associated with a past common-envelope episode. A subset of these binaries host unusually massive WDs in the $1-1.3$\,\msun\ range \citep{Yamaguchi_2024_WidePCEBs}, although such objects remain rare compared to the general field population \citep{Hallakoun_2024}.

Standard stellar evolution models link WD mass to progenitor mass through the initial-to-final mass relation (IFMR). For instance, in isolation, a $1.2$\,\msun\ WD implies a progenitor of approximately $7$\,\msun\ \citep[e.g.,][]{Cunningham_2024}. While mass loss and transfer in close binaries can modify this mapping \citep[e.g.,][]{Ironi_2025, Shahaf_2025}, the fundamental prerequisite persists: the formation of a massive WD generally necessitates a massive stellar progenitor. 

However, hierarchical triples introduce a distinct evolutionary pathway. In these systems, secular evolution and mass loss can drive the inner binary to coalescence, resulting in a single WD whose mass is decoupled from that of a single progenitor \citep{Perets_2012, Shariat_2023}. This channel can populate both the low- and high-mass extremes of the WD mass distribution \citep[e.g.,][]{Shariat_2025, Zhang_2026}. Recent population-synthesis works clearly imply that inner-binary mergers frequently produce long-lived post-merger binaries, systems where a massive WD that is the merger product remains bound to the former tertiary companion \citep{Rajamuthukumar_2023, Shariat_2025, Shariat_2026}. Subsequently, some of these binaries may undergo a second merger, providing a pathway to Type~Ia supernovae (SNe~Ia). The present work aims to establish observational diagnostics and evidence for identifying systems currently in this intermediate evolutionary phase.

Distinguishing whether a massive WD formed via single-star evolution or a merger event is challenging, as observations are limited to the final system configuration. However, chemical enrichment offers a powerful diagnostic tool. Barium (Ba) enhancement in a companion star signifies accretion from an Asymptotic Giant Branch (AGB) donor that underwent third dredge-up and \textit{s}-process nucleosynthesis. Crucially, AGB stars more massive than ${\sim}\,4$\,\msun\ are inefficient at producing or dredging up \textit{s}-process elements \citep[e.g.,][]{Karakas_2016, Karakas_2018, Rekhi_2024, Rekhi_2026, Yamaguchi_2025_Barium}. Consequently, massive WDs with Ba-enhanced companions are unlikely to have descended from a single massive progenitor. A plausible interpretation is that such WDs formed through the merger of two lower-mass stars, at least one of which was capable of supplying the observed \textit{s}-process material.

The role of triple interactions in driving chemical enrichment is now well established \citep{Gao2023}. Similarly, recent theoretical studies have highlighted the significant contribution of these systems to WD mergers, which may ultimately yield SNe~Ia. Specifically, \citet{Rajamuthukumar_2023} and \citet{Shariat_2026} showed that the ``double-merger'' channel, wherein an inner-binary merger produces a massive WD that subsequently coalesces with the tertiary, can account for a substantial fraction of the observed SN~Ia rate. Furthermore, specific signatures of coalescence in both binary and triple configurations have been predicted \citep{Kummer2025, Vynatheya2025}. Despite these advances, chemical abundance analysis remains an underutilized diagnostic for constraining the complex dynamical history of these systems.

This work presents an empirical methodology for identifying merger-product WDs via their chemical enrichment signatures. We benchmark this approach against the well-known system IK~Pegasi \citep[IK~Peg;][]{Smalley_1996}. While the precise evolutionary history of IK~Peg remains debated, it serves as a critical prototype, illustrating how chemical and orbital constraints can be synthesized to inform future searches. We subsequently apply this framework to a select sample of binaries with existing abundance measurements. Ultimately, this study establishes a foundation for targeted observational campaigns designed to constrain the incidence of such systems and elucidate their role in binary and hierarchical evolution, including pathways leading to SNe~Ia.

\section{IK~Peg}

IK~Pegasi (IK~Peg; also known as HR\,8210 or HD\,204188), a 6th-magnitude star located about 46\,pc from the Sun, was first cataloged in the Bonner Durchmusterung catalog \citep{Argelander_1903}. It was subsequently classified as an A-type star in the Revised Harvard Photometry catalog \citep{Pickering_1908}. In 1923, the system was identified as a binary upon the detection of radial-velocity (RV) variations in its absorption lines. These variations indicated a circular orbit with a period of $\sim 22$\,d \citep[][see Appendix~\ref{app:IKPeg} for a more detailed description of the system]{Harper_1927, Harper_1935}.

The system, composed of a $1.2$~\msun\ WD and a chemically peculiar pulsating A-type MS star, presents an evolutionary anomaly: its $0.2$\,au orbital separation is difficult to reconcile with standard models of binary interaction. It is generally assumed that the WD's progenitor was a massive star with a large radius that would have exceeded the current orbital separation. This implies that a common-envelope phase has occurred \citep[e.g.,][]{Wonnacott_1993}. Such strong interaction between the components is typically expected to lead to an inspiral, shrinking the orbit to only a few solar radii (e.g., \citealt{Toonen_2013}; although recently \citet{Belloni_2024} suggested that inefficient common-envelope evolution can result in wider orbits).

\begin{figure*}
    \centering
    \includegraphics[width=\linewidth]{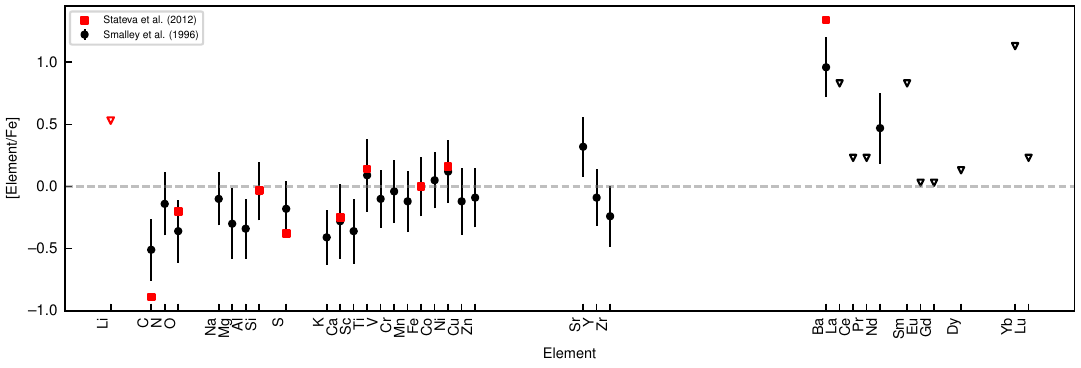}
    \caption{IK~Peg\,A element abundance relative to iron, from \citet[][black circles; $\text{[Fe/H]}=+0.17\pm0.17$]{Smalley_1996}, and from \citet[][red squares; $\text{[Fe/H]}=+0.07$]{Stateva_2012}. Open triangles mark upper limits.}
    \label{fig:IKPegAbundances}
\end{figure*}

The MS companion in IK~Peg displays a distinct chemical anomaly, defined by moderate carbon depletion and a pronounced barium overabundance (Figure~\ref{fig:IKPegAbundances}). While early investigations attributed these peculiarities to intrinsic stellar processes \citep{Cowley_1969, Abt_1969}, subsequent analyses have identified mass transfer from the WD progenitor as the likely source of the pollution \citep[][see also the detailed discussion in Appendix~\ref{app:IKPeg}]{Landsman_1993, Smalley_1996}. 

Observable enrichment in the companion requires that the WD progenitor was an efficient producer of \textit{s}-process elements. These yields depend primarily on metallicity and mass. Observations show an anti-correlation between AGB \textit{s}-process abundances and metallicity \citep{Roriz_2021, Vilagos_2024}, while theoretical models predict the highest yields for initial masses of about $2-4$\,\msun\ \citep{Karakas_2016, Karakas_2018}. The joint dependence of these yields on mass and metallicity has been confirmed empirically by \citet{Rekhi_2024, Rekhi_2026}. They showed that \textit{s}-process enrichment in the MS companions is found predominantly in sub-solar metallicity systems with WD masses of $0.6-0.75$\,\msun, consistent with progenitors of $2$–$4$\,\msun. For systems hosting more massive WDs, enrichment is detected only at significantly lower metallicities, [Fe/H] $\lesssim -0.4$ (see Figure~\ref{fig:Mwd_FeH_BaFe}).

The evolutionary history of the system is constrained by three factors: age, chemical yields, and the WD mass. The upper age limit of 600\,Myr for IK~Peg\,A \citep{Wonnacott_1994} requires that the progenitor of the inner primary was $\gtrsim 3$\,\msun. To account for the observed Ba enhancement, this progenitor must have had a mass of $2$--$4$\,\msun\ so that it could produce sufficient \textit{s}-process material during the AGB phase \citep{Karakas_2016}. In contrast, the WD mass of ${\sim}\,1.2$\,\msun\ implies a single-star progenitor with an initial mass $\gtrsim 6$\,\msun\ \citep[e.g.,][]{Cunningham_2024}. Binary interaction would further reduce the remnant mass \citep{Ironi_2025, Shahaf_2025}, which would require an even more massive progenitor. 

The combined constraints on age, \textit{s}-process yields, and WD mass can be reconciled by invoking a merger. As an illustrative example, a coalescence in the inner binary with a total mass of at least $\sim 6$\,\msun\ can produce the observed WD while allowing one component to lie in the $2-4$\,\msun\ regime required for sufficient Ba production. In this scenario, the tertiary, now the observed primary, likely accreted only a small amount of material, enriching its surface without significantly changing its mass.

\section{IK~Peg--Type Systems}
\subsection{Definition and Identification}

For decades, IK~Peg was the only known post-mass transfer system hosting a massive ($\gtrsim 1$\,\msun) WD and an orbital period on the order of tens of days \citep[e.g.][]{Parsons_2023}. Recently, \citet{Yamaguchi_2024_WidePCEBs} reported five additional such systems, found in the Gaia's non-single star (NSS) catalog \citep{Gaia_Arenou_2023}. However, binary systems with massive WD companions and chemical abundance or age inconsistencies exist also with orbital separations of hundreds of days. Motivated by the anomalies observed in IK~Peg, we propose that \textit{s}-process pollution serves as a valuable tracer for identifying post-merger binaries. We suggest the following definition of an \textbf{IK~Peg--type system}, as an \emph{\textit{s}-process enriched system hosting a massive $\gtrsim$ 0.8\,\msun\ WD, with an orbital separation of 0.1--10\,au}.

Single-star evolution constrains the WD progenitor mass via the IFMR and sets an upper limit on the \textit{s}-process enrichment achievable during the AGB phase at a given metallicity. Conservatively, systems incompatible with single-progenitor evolution satisfy
\begin{equation}\label{eq: ba cut}
\text{[Ba/Fe]}_\text{obs} \gtrsim \text{[Ba/Fe]}_\text{max}^\textsc{agb},
\end{equation}
where the left-hand side is the Ba abundance measured in the WD companion's atmosphere, and the right-hand side represents the maximum enrichment from any single AGB progenitor consistent with the WD mass.

Figure~\ref{fig:BaFeVsMwd} displays the maximal theoretical Ba yields across different metallicities and progenitor masses \citep{Karakas_2016, Karakas_2018} as a function of the expected WD mass, assuming a single-star IFMR \citep{Cunningham_2024}.
Since any realistic mass transfer or dilution process can only reduce the observable surface enrichment, systems violating this condition are unlikely to descend from single-progenitor evolution under any plausible assumptions, provided the companion itself is not an AGB star.

\begin{figure}
    \centering
    \includegraphics[width=\columnwidth]{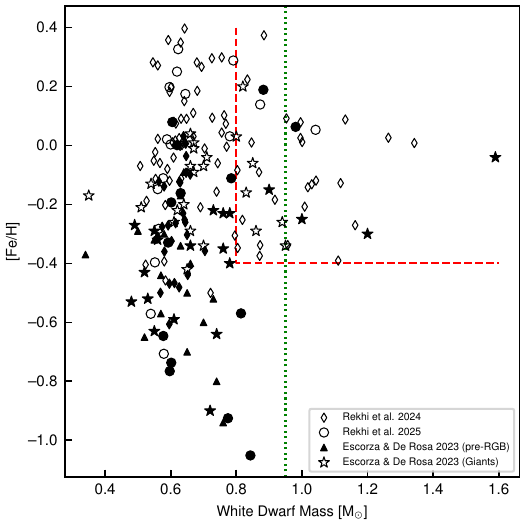}
    \caption{The relationship between WD mass, companion metallicity, and Ba abundance. Gaia MS+WD systems with measured Ba abundances are plotted as diamonds \citep{Rekhi_2024} and circles \citep{Rekhi_2026}. Known Ba dwarfs (pre-red giant branch; pre-RGB) hosting WD companions from \citet{Escorza_2023} are shown as triangles, while star symbols denote Ba giants. Filled symbols indicate Ba-enriched systems, defined as $\text{[Ba/Fe]} > +0.25$ for the \citet{Rekhi_2024, Rekhi_2026} samples, or classified as Ba dwarfs/strong Ba giants in the \citet{Escorza_2023} sample. The red dashed line delineates the tentative empirical limit beyond which barium enrichment is inconsistent with a single-progenitor scenario. The green dotted line marks the most conservative limit, based on single-progenitor evolution models (see Figure~\ref{fig:BaFeVsMwd}).}
    \label{fig:Mwd_FeH_BaFe}
\end{figure}

\begin{figure}
    \centering
    \includegraphics[width=\columnwidth]{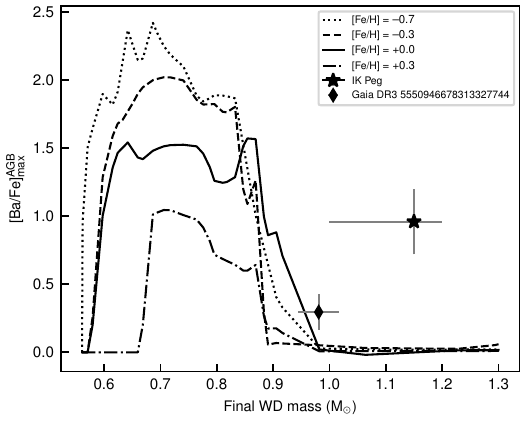}
    \caption{Maximal [Ba/Fe] abundance attained during the AGB phase for metallicities $\text{[Fe/H]} \approx -0.7$ \citep{Karakas_2018}, $-0.3$, $0$, and $+0.3$ \citep{Karakas_2016}, plotted against the final WD mass of a single progenitor using the \citet{Cunningham_2024} IFMR. For comparison, the positions of IK~Peg (star) and the IK~Peg--type candidate Gaia~DR3\,5550946678313327744 (diamond; \citealt{Rekhi_2026}) are indicated. Both systems have near-Solar metallicity.}
    \label{fig:BaFeVsMwd}
\end{figure}

Based on the theoretical predictions in Figure~\ref{fig:BaFeVsMwd}, it is reasonable to assume that detectable Ba enrichment is not expected for companions of WDs more massive than $\gtrsim 0.95$\,\msun, under standard single-progenitor evolution. This conservative domain is marked by the green dotted line in Figure~\ref{fig:Mwd_FeH_BaFe}. However, IK~Peg--type systems have likely experienced enhanced mass loss during past interaction episodes. As recently demonstrated by \citet{Ironi_2025}, post-mass transfer WDs are typically at least $0.1$\,\msun\ less massive than predicted by the single-star IFMR. When combined with the expected dilution of Ba in the companion's photosphere, the theoretical limit of $\gtrsim 0.95$\,\msun\ appears overly conservative. We therefore draw a more realistic, empirical boundary based on the observed sample of \citet{Rekhi_2024, Rekhi_2026}, tentatively placing this limit at WD masses $>0.8$\,\msun\ and metallicities $\text{[Fe/H]} > -0.4$ (red dashed line in Figure~\ref{fig:IKPegAbundances}). We emphasize that this empirical boundary is currently derived from a limited sample, which does not fully populate the parameter space (as illustrated in Figure~\ref{fig:Mwd_FeH_BaFe}). Dedicated observational efforts, such as the recent survey by \citet{Yamaguchi_2025_Barium}, are expected to provide improved constraints on this limit.

A further challenge in identifying IK~Peg--type systems lies in the inherent uncertainty of abundance estimates. Spectral fitting is highly sensitive to the choice of spectral regions, the underlying radiative transfer code, the adopted zero-points (solar abundance scale), and the fitting methodology. These dependencies introduce significant systematic uncertainties \citep[e.g.,][]{BlancoCuaresma_2019}. However, large-scale population studies provide a means to characterize and calibrate these biases. For instance, benchmarking the \textsc{pysme}-based pipeline of \citet{Rekhi_2026} against GALAH DR3 estimates \citep{Buder_2021} reveals a systematic offset of $\lesssim0.1$\,dex (see Appendix~\ref{app:BaAbundnace}, and see also figure~7 in \citealt{Yamaguchi_2025_Barium}). We thus adopt an enrichment threshold of $0.25$\,dex, and define IK~Peg--type binaries as systems that exceed this threshold despite residing in a region of parameter space where theoretical AGB models predict negligible enrichment ($\text{[Ba/Fe]}_\text{max}^\textsc{agb} \simeq 0$). Consequently, systematic abundance surveys can, at a minimum, distinguish relative enrichment levels and anchor the absolute scale using control samples with independently constrained abundances. Such controls include, for example, MS+WD binaries with massive WD companions that exhibit no \textit{s}-process enrichment, or single MS stars with comparable mass, metallicity, and magnitude.

\begin{figure}
    \centering
    \includegraphics[width=\columnwidth]{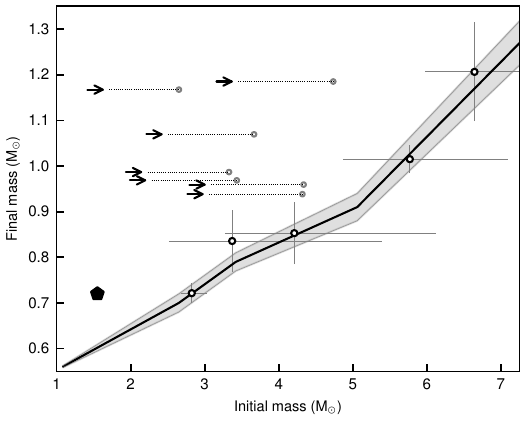}
    \caption{Comparison of initial and final masses adapted from \citet{Ironi_2025}. Arrows and gray dots mark the cluster turn-off mass and the median progenitor mass (assuming a \citeauthor{Kroupa_2001} IMF), respectively. Open circles indicate systems consistent with the \citet{Cunningham_2024} IFMR (solid line; $1\sigma$ uncertainty in gray). The pentagon marks the estimated position of the Blue Lurker WOCS~14020 \citep{Leiner_2025}.}
    \label{fig:IFMR}
\end{figure}

Although the \citet{Rekhi_2026} sample includes a few \textit{s}-process enriched IK~Peg--type candidates in the region where no enrichment is expected, a definitive confirmation requires a direct constraint on the progenitor mass. Ideally, this would be achieved by identifying a tension between the WD mass, its cooling age, and the total system age---a quantity that is typically difficult to determine. A promising alternative is to target binaries within open clusters, where the cluster age provides an independent constraint on the progenitor mass. A particularly important case is WOCS~14020, identified by \citet{Leiner_2025}. This ``blue lurker'' exhibits evidence of past mass transfer through its rapid rotation rather than chemical enrichment. By comparing the system's age with the WD cooling age, the authors concluded that the most likely formation pathway involves a merger within a hierarchical triple system. In this scenario, the interaction truncates the AGB phase, thereby precluding Ba enrichment. \citet{Ironi_2025} identified a dozen binaries hosting massive WD companions in sufficiently old open clusters, where systems similar to WOCS~14020 could be identified (see Figure~\ref{fig:IFMR}). However, the current sample size is limited, and reliable age estimates necessitate space-based ultraviolet data, which are unavailable for all sources.

A key advantage of IK~Peg--type systems is that they extend this framework from clusters to the field, enabling scalable population-level searches for post-merger binaries. To the best of our knowledge, all IK~Peg--type systems currently discussed in the literature should be considered candidate post-merger binaries. Once this connection is more robustly established, through definitive detections like WOCS~14020 and the analysis of larger samples, Ba enrichment may transition from a supporting indicator to direct evidence of this evolutionary channel. Achieving this requires systematic follow-up of high-probability candidates.

\subsection{Expected Occurrence Rate}
\label{sec:occurrence}

We estimate the expected number of IK~Peg--type systems in the Gaia sample using two independent approaches: a direct empirical constraint and an extrapolation from wider post-merger populations.

Gaia has identified approximately $5000$ MS$+$WD candidates, of which about $3000$ are robustly classified \citep{Shahaf_2024}. Using a volume-complete subsample, \citet{Hallakoun_2024} found that only ${\sim}\,3\%$ of WDs more massive than ${\sim}\,0.7$\,\msun\ (of any kind) host K- or M-dwarf companions. A complementary analysis by \citet{Rekhi_2024} measured Ba abundances for 102 systems drawn from the same population by cross-matching it with the GALAH catalog, 25 of which host WDs more massive than $0.8$\,\msun\ and exhibit metallicities larger than $-0.4$. None showed evidence for an enriched companion (see Figure~\ref{fig:Mwd_FeH_BaFe}).\footnote{While the FEROS sample of \citet{Rekhi_2026} identifies a few IK~Peg--type candidates (discussed below; see Figure~\ref{fig:Mwd_FeH_BaFe}), it suffers from strong selection effects. Consequently, this sample cannot be used to reliably quantify the intrinsic frequency of such systems.} This non-detection sets a $95\%$-confidence upper limit of $\approx 11\%$ on the fraction of IK~Peg--type systems among MS$+$WD binaries hosting massive WDs.

An independent estimate comes from the wider post-merger population. \citet{Heintz2022} analyzed wide double WDs identified by Gaia and showed that in roughly one-third of the systems the more massive WD is paradoxically the younger component. Such a configuration is incompatible with any monotonic IFMR, and they interpreted this discrepant population as consisting of merger remnants or unresolved triples. These systems, however, occupy separations much larger than those considered in this work. Population-synthesis models by \citet{Shariat_2025} predict that inner-binary mergers in solar-type triples frequently produce long-lived post-merger binaries, but that only ${\sim}\,2\%$ of these remnants occupy separations of $0.1$--$10$\,au; the vast majority reside at much wider orbits, comparable to those studied by \citet{Heintz2022}. 

If the wide post-merger population identified by \citet{Heintz2022} traces the same underlying channel as the closer systems considered here (an assumption that remains to be tested), then the \citet{Shariat_2025} scaling implies that ${\sim}\,0.5\%$ of Gaia MS$+$WD binaries should host a post-merger remnant in the IK~Peg--type separation regime. This corresponds to roughly ${\sim}\,25$ systems in the current sample, a subset of which should exhibit detectable \textit{s}-process enhancement. This estimate is broadly consistent with the empirical upper limit derived above, suggesting that the current Gaia sample may contain a few dozen IK~Peg--type systems awaiting identification.

\section{Possible formation histories}

The theoretical framework for producing massive WDs through inner-binary mergers in hierarchical triples has been developed in detail by \citet{Rajamuthukumar_2023, Shariat_2023, Shariat_2025, Kummer2025}. Here we summarize the key elements of their evolutionary scenario to contextualize the chemical enrichment signatures proposed in this work. Our objective is not to present an exhaustive model, but to demonstrate how post-merger binaries naturally account for the observed properties of IK~Peg--type systems: the high WD mass, the intermediate orbital separation, and the companion's chemical peculiarities. 

\subsection{Dynamical Origin of the Present-day Binary}

To be specific, we adopt the evolutionary sequence of \citet[see their figure~2]{Shariat_2025} as a representative test case. In their simulations, a merger of an intermediate-mass inner binary naturally produces a massive WD remnant, providing a physically motivated basis for discussion.

The system begins as a stable hierarchical triple. The inner binary comprises an intermediate-mass primary (${\sim}\,4.5$\,\msun) and a ${\sim}\,1$\,\msun\ secondary with a separation of a few au. This inner pair is orbited by an A-type tertiary (${\sim}\,2$\,\msun) at a distance of ${\sim}\,60$\,au. Both the inner and outer orbits are assigned initial eccentricities of $e \sim 0.5$, consistent with the broad distribution observed in intermediate-mass triple systems.

As the inner primary ascends the AGB, its radial expansion eventually fills its Roche lobe, triggering unstable mass transfer onto the companion. This interaction precipitates a common-envelope phase, which effectively strips the donor's envelope and drives orbital decay until the inner components coalesce. The resulting merger product is an over-massive, rapidly evolving giant-like object that retains the chemically processed material generated during the AGB phase.

If the mass ejected during the merger is sufficiently small, the outer orbit remains gravitationally bound, effectively converting the system into a post-merger binary. For a brief interval, tidal interactions with the extended merger product cause the orbit of the former tertiary to shrink and circularize. Following a short evolutionary phase, the merger remnant sheds its remaining envelope and contracts to form a WD. Consequently, significant orbital evolution arrests, leaving a stable configuration comprising a massive WD and the former tertiary star at a separation of a few au.

\subsection{Origin of the Tertiary's Chemical Enrichment}

Due to their thin convective envelopes, A-type MS stars act as sensitive probes for surface pollution. The limited mass of this convective zone ensures that accreted material undergoes minimal dilution, allowing the photospheric composition to preserve signatures of even brief \textit{s}-process accretion episodes.

Such accretion events may proceed via two distinct channels during the merger sequence. First, as the inner primary ascends the thermally pulsating AGB (TP-AGB), it drives a slow wind laden with chemically enriched material, a fraction of which is intercepted by the outer tertiary. We provide an estimate of this accreted mass below. A second enrichment opportunity arises during the subsequent common-envelope phase of the inner binary, when a significant portion of the donor's envelope is expelled from the system and potentially captured by the tertiary. Furthermore, the common-envelope interaction itself may facilitate the dredge-up of \textit{s}-process elements, potentially occurring even in the absence of thermal pulses.

Typical AGB stars drive slow winds with characteristic velocities of order $3$--$30$\,\kms\ \citep{Hofner2018}. Applying a standard Bondi-Hoyle-Lyttleton approximation demonstrates that detectable chemical enrichment is plausible within this regime. The accretion efficiency, defined as the ratio between the accreted mass and the mass lost by the donor, is
\begin{equation}
    f_{\rm acc} \approx 0.05  
    \left(\frac{M_{\rm a}}{2\,M_\odot}\right)^2
    \left(\frac{v}{20\,{\rm km\,s^{-1}}}\right)^{-4}
    \left(\frac{d_{\rm p}}{20\,{\rm AU}}\right)^{-2},
\end{equation}
where $M_\text{a}$ is the mass of the A-type accretor, $d_\text{p}$ is the periastron separation of the outer orbit, and $v$ is the characteristic wind velocity.
Strictly speaking, this approximation depends on the relative velocity between the wind and the accretor. However, in the configurations considered here, the wind and orbital velocities are likely of comparable magnitude; thus, considering only the wind velocity introduces a correction of order unity. Under this approximation, the expression implies that a percent-level fraction of the AGB outflow can be intercepted by the A-type companion during periastron passage.

During the final stages of its evolution, the AGB donor typically ejects of order $0.5$\,\msun\ \citep{Hofner2018}. However, only a fraction of this material is available for capture by the tertiary. While the instantaneous accretion efficiency can reach a few percent, the tertiary resides in the effective accretion regime for only a brief portion of its orbit, roughly $1$--$10\%$ of the cycle. Consequently, the net accretion efficiency is reduced to the level of $10^{-3}$ to $10^{-2}$. This implies a total accreted mass of
\begin{equation}
    M_\text{acc} \lesssim 10^{-3}\,\text{\msun}.
\end{equation}
The accreted material is incorporated into the very thin convective envelope of the A-type accretor \citep{Turcotte_1998}. Consequently, the dilution is minimal, preserving a potentially detectable \textit{s}-process enhancement in the photosphere. Still, this estimate is intended only to demonstrate plausibility; the efficiency of wind accretion and enrichment retention is expected to vary widely between systems.

\section{Discussion}

At present, the connection between post-merger systems and IK~Peg--type binaries remains circumstantial, and relies on our understanding of slow neutron–capture nucleosynthesis. Nevertheless, recent large-scale astrometric and spectroscopic surveys have uncovered a growing population of MS$+$WD binaries exhibiting chemical patterns broadly consisting with this scenario. Table~\ref{tab:ikpeglike} in Appendix~\ref{app:IKPegType} provides a summary of such candidates reported in the recent literature. In the following subsections, we discuss several representative examples and outline observational strategies to validate their evolutionary history.

\subsection{IK~Peg--type Candidates}
\label{sec:candidates}

In this section, we briefly discuss several prominent candidates identified in the recent literature.

The systems HD\,49641, HD\,31487, HD\,92626, and HD\,44896 \citep{Escorza_2023} are well-established classical Ba giants that serve as valuable comparative benchmarks for IK~Peg--type candidates. All are evolved red giants residing in long-period binaries, displaying strong \textit{s}-process enhancements attributed to historical mass transfer from a companion that has since evolved into a WD. Their orbital periods and eccentricities are consistent with wind-driven accretion in wide systems. Although these objects represent a later evolutionary phase than the MS$+$WD systems considered here, they share the same post--mass-transfer origin, thereby providing empirical benchmarks for the chemical signatures of AGB pollution. Consequently, they can be viewed as the evolved counterparts of IK~Peg--type systems, linking the population of chemically peculiar giants to their unevolved progenitors.

A recent, representative example with a polluted MS star is Gaia~DR3\,5550946678313327744, identified by \citet{Rekhi_2026}. This binary is characterized by an orbital period of ${\sim}\,800$\,d and a low eccentricity of ${\sim}\,0.05$, hosting a ${\sim}\,0.98$\,\msun\ WD alongside a solar-metallicity, Sun-like companion. While the observed Ba enhancement of $\text{[Ba/Fe]} \simeq 0.3$ appears modest in absolute terms, it lies well above the theoretical upper limit for single-progenitor AGB yields at this specific WD mass (see Figure~\ref{fig:BaFeVsMwd}). This discrepancy becomes even more pronounced when accounting for the effects of enhanced mass loss and dilution.

Two interesting systems were recently identified by \citet{Yamaguchi_2025_Barium}: The first, Gaia~DR3\,1386979565629462912, is characterized by a highly eccentric long-period orbit (${\sim}\,870$\,d) and exhibits strong Ba enrichment ($\text{[Ba/Fe]} \simeq 1.3$), with an inferred dark companion mass of $1.6 \pm 0.2$\,\msun. The second, Gaia~DR3\,5648541293198448512, similarly hosts a dark companion ($1.5 \pm 0.3$\,\msun) on a highly eccentric wide orbit, yet displays only modest enrichment ($\text{[Ba/Fe]} \simeq 0.26$). While both inferred masses formally exceed the Chandrasekhar limit, the underlying astrometric solutions await independent validation. At these mass ranges, metallicity is not a governing constraint. Should these high masses be confirmed, the companions are likely non-WD objects (i.e., neutron stars). Conversely, a downward revision placing them into the massive-WD regime would render these systems prime candidates for follow-up and detailed modeling, given their distinct enrichment signatures.

While these candidates offer significant potential, definitive validation demands more than just refined orbital parameters; it necessitates the development of supplementary observational diagnostics to distinguish merger remnants from the products of standard single-star evolution.

\subsection{Validation of the Formation Channel in Open Clusters}

MS$+$WD binaries residing within open clusters offer a robust and direct validation of the post-merger scenario. The age of the parent cluster sets a strict lower bound on the progenitor mass via the MS turn-off. Direct observations of the WD yields its cooling age; subtracting this from the total cluster age isolates the MS lifetime, and thus the mass, of a single progenitor. Consequently, any system where the progenitor mass derived from these timing constraints is inconsistent with the observed WD mass (via the IFMR) mandates a non-standard evolutionary history. This provides confirmation of a merger origin independent of chemical abundance signatures \citep[e.g., WOCS~14020; see][]{Leiner_2025}.

Recent investigations have revealed a small population of massive WD binaries within open clusters \citep[e.g.,][]{Ironi_2025}. In several cases, the cluster turn-off mass is sufficiently low that the observed WD masses are unlikely to reconcile with single-star evolution, suggesting a merger origin (see Figure~\ref{fig:IFMR}). Chemical abundance measurements serve as an efficient secondary filter: the presence of IK~Peg--type chemical signatures among cluster members identifies the most promising candidates. Subsequently, targeted ultraviolet observations of this refined subsample enable the precise determination of WD cooling ages. By combining these cooling ages with the cluster's total age, one can derive the required progenitor mass under single-star assumptions. This strategy provides a minimally model-dependent test of the post-merger origin for \textit{s}-process enrichment, utilizing Ba abundance as a tracer.

\subsection{Rate Estimates from Well-defined Samples}

While individual systems can demonstrate feasibility, constraining the incidence of IK~Peg--type binaries requires the homogeneous follow-up of a well-defined sample. Gaia astrometry has revealed thousands of MS$+$WD candidates, including a substantial sub-population hosting massive WDs. For these systems, medium-resolution optical spectroscopy is sufficient to detect \textit{s}-process enrichment at levels exceeding the theoretical maximum for single-progenitor evolution at a given WD mass. A targeted spectroscopic campaign focusing on binaries with massive WDs, employing uniform selection and analysis criteria, enables a direct measurement of the fraction of systems inconsistent with standard binary evolution. Such a sample would provide a robust estimate of the IK~Peg--type occurrence rate in the field, enabling a systematic exploration of its dependence on WD mass, orbital separation, and eccentricity. This strategy builds upon existing high-resolution efforts (using FEROS), and is readily scalable to larger samples.

Figure~\ref{fig:Mwd_FeH_BaFe} presents the current population of binaries with massive WDs and measured Ba abundances from \citet{Escorza_2023} and \citet{Rekhi_2024, Rekhi_2026}. However, its selection has been spurious, or inconsistent as it is complied by several different campaigns. Considering only the few IK~Peg--type candidates to MS stars were identified, this is consistent with occurrence rate ${\lesssim}\,10\%$, assuming selection effects are negligible  (in agreement with the estimate shown in Section~\ref{sec:occurrence}). Further study is needed to set, at the very least, a realistic and robust lower limit on this rate.

\section{Implications to Type~Ia Supernovae}

\subsection{The Double-merger Pathway to Type~Ia Supernovae}

SNe~Ia, the thermonuclear explosions of WDs, are primary nucleosynthetic sources of heavy elements and serve as ``standard candles'' for cosmic distance measurements. 
These events underpinned the discovery of the accelerating expansion of the Universe and dark energy. Yet, despite their fundamental importance, the nature of their progenitor systems remains elusive. While a binary origin involving a WD is well-established, the nature of the companion star and the specific physical mechanism triggering the explosion remain subjects of active debate \citep[see e.g.,][for reviews]{Maoz_2014, Liu_2023, Ruiter_2025}.

The currently favored progenitor scenario, known as the ``double degenerate'' channel \citep{Tutukov_1981, Iben_1984, Webbink_1984}, involves a binary system comprising two WDs. In this model, the orbit decays due to the emission of gravitational waves, causing the system to lose energy and angular momentum until the components merge and potentially detonate. However, the ``double degenerate'' channel itself encompasses a broad spectrum of scenarios, distinguished by the physical properties of the system, the specific detonation mechanism, and the observational aftermath of the explosion \citep[e.g.,][]{Ruiter_2025}.

If the double-degenerate channel is the primary source of normal SNe~Ia in Milky Way-like galaxies, the Galactic double-WD merger rate must, at a minimum, equal the observed SN~Ia rate (since not all double WDs mergers explode). Based on large spectroscopic WD samples, \citet{Maoz_2018} estimated that the merger rate in the local neighborhood exceeds the SN~Ia rate by a factor of a few \citep[see also][]{Badenes_2012, Maoz_2012, Maoz_2017}.

Nevertheless, the fraction of mergers capable of triggering a thermonuclear explosion remains unconstrained. Simulations suggest that reproducing the observational signatures of typical SNe~Ia requires a WD mass of $\gtrsim 0.9$\,\msun\ \citep{Shen_2018, Shen_2021}. Yet, such massive WDs are intrinsically rare. If a significant fraction of massive WDs are indeed merger products (see Section~\ref{sec:intro}), and if SNe~Ia require a massive WD within a binary system, this implies a hierarchical triple origin, as previously noted by \citet{Maoz_2018}. In this ``double merger'' scenario, the inner binary merges to create the requisite massive WD, which subsequently merges with the tertiary companion. 

The outcome of this channel depends sensitively on the primordial properties of triple systems, including their separation, mass-ratio, and eccentricity distributions, as well as the overall multiplicity fraction. The population-synthesis models of \citet{Rajamuthukumar_2023} and \citet{Shariat_2026} demonstrate that hierarchical triples can efficiently produce massive WDs and enable subsequent thermonuclear explosions through this double-merger pathway. Their simulations indicate that this channel can contribute meaningfully to the observed SN~Ia rate within current theoretical uncertainties. The IK~Peg--type systems identified in this work represent candidate intermediate stages of this evolutionary sequence, offering an empirical window onto a channel that has thus far been characterized only theoretically.

\subsection{IK~Peg and the SNe Ia Rate}

We emphasize that the following estimates should be interpreted as conservative upper limits and illustrative contexts, rather than as strict physical constraints.

\citet{Yamaguchi_2025_PopDemographics} estimated that ${\sim}\,0.4\%$ of Sun-like stars host WD companions at separations of order 1\,au, while self-lensing constraints permit a slightly higher fraction of up to ${\sim}\,1.1\%$ \citep{Yamaguchi_2024_TESS}. As shown in Section~\ref{sec:occurrence}, massive WDs constitute a few percent of the MS$+$WD population in this separation regime. Consequently, only a few ${\sim}\,1\times10^{-4}$ of Sun-like stars are expected to harbor a massive WD companion at ${\sim}1$\,au. Extrapolating this value across the $0.1$--$10$\,au range, under the assumption of a log-flat orbital distribution, implies an integrated occurrence rate of $\lesssim 10^{-3}$.

At their current orbital separations, a non-negligible fraction of these systems are expected to undergo a second merger within the evolutionary lifetime of the MS companion (the former tertiary). Adopting a characteristic evolutionary timescale of ${\sim}\,10$\,Gyr for a Sun-like star, and assuming that these post-merger systems descend from hierarchical triples with a total initial mass of ${\sim}\,5$\,\msun, we estimate a specific double-merger rate of up to ${\sim}\,2\times10^{-14}$\,yr$^{-1}$\,\msun$^{-1}$.

As far as we know, empirical constraints on systems with more massive primordial tertiaries are currently even sparser than for the solar-type population discussed above. In general, these systems are characterized by evolutionary lifetimes roughly an order of magnitude shorter and total masses larger by a factor of a few. Given the higher multiplicity fraction of more massive stars, we postulate that the merger rate for IK~Peg--type systems hosting A-type primaries is comparable to, or potentially higher than, our estimate for Sun-like stars. Notably, this estimated rate constitutes a non-negligible fraction of the specific SN~Ia rate for a Milky Way--mass Sbc galaxy \citep[$\approx1.1\times10^{-13}$\,yr$^{-1}$\,\msun$^{-1}$;][]{Li_2011}. These estimates suggest that post-merger binaries within the separation range accessible to Gaia may contribute significantly to the SN~Ia rate.

These order-of-magnitude rate estimates should be interpreted as plausible upper limits. IK~Peg--type systems offer an empirical, conservative probe of the double-merger channel. Adopting the hypothesis that these systems descend from inner-binary mergers, their measured incidence establishes a robust lower limit on the total contribution of this pathway to the SN~Ia rate. This follows from the fact that not all primordial tertiaries acquire detectable chemical enrichment; thus, the enriched sample represents a strict minimum baseline, independent of detailed delay-time models.

\section{Outlook}

Chemical enrichment serves as a fossil record of a system's formation history, preserving evidence of past mass transfer and merger events long after the associated dynamical signatures have dissipated. In the context of MS$+$WD binaries, \textit{s}-process enrichment provides a vital diagnostic tracer; it directly links the observed surface composition to the nucleosynthetic yields of the progenitor, thereby breaking the degeneracy between competing evolutionary pathways.

The combination of Gaia astrometry with large-scale spectroscopic surveys enables the systematic application of this diagnostic at the population level. Upcoming Gaia data releases, particularly Gaia DR4, will substantially expand the sample of astrometric binaries with well-characterized orbital solutions, providing a well-defined target list for homogeneous abundance follow-up. Targeted spectroscopic surveys of these systems will allow the incidence of IK~Peg--type binaries to be measured robustly and their dependence on WD mass, orbital separation, and metallicity to be explored.

More broadly, the use of chemical enrichment as a tracer of formation history extends well beyond the specific case of MS$+$WD binaries. Recent discoveries of lithium in Gaia-identified neutron stars underscore the potential of surface abundances to retain signatures of exotic evolutionary channels \citep{El-Badry2024_lithium}. Analogously, should some IK~Peg--type candidates prove to host neutron stars rather than WDs, their Ba enhancement may point to a history of mass transfer preceding an accretion-induced collapse. Examples for such cases are the massive, enriched systems reported by \citet[][see Section~\ref{sec:candidates}]{Yamaguchi_2025_Barium}, or some of the more massive candidates identified in \citet{Escorza_2023}, \citet{Shahaf_2023, Shahaf_2024} and \citet{Yamaguchi_2024_WidePCEBs}. In this scenario, the detailed feasibility of which remains to be assessed, the \textit{s}-process enrichment predates the collapse and reflects the progenitor's evolutionary history rather than the collapse event itself. Thus, chemical tagging emerges as a scalable and powerful tool for reconstructing the evolutionary pathways of compact-object binaries across the mass spectrum.

%% Please use the acknowledgment and contribution environments. This will 
%% be anonomyized when the "anonymous" style option is used. 
\begin{acknowledgments}
We thank Jiadong Li, Dan Maoz, Tsevi Mazeh, Johanna M\"uller-Horn, Ben Pennell, Silvia Toonen, Jaime Villase\~nor, and the anonymous referee for helpful discussions.

N.H. acknowledges support from the Planning \& Budgeting Committee of the Israeli Council for Higher Education. S.S. and H.W.R. acknowledge support from the European Research Council for the ERC Advanced Grant (101054731). This research was supported in part by grant NSF PHY-2309135 to the Kavli Institute for Theoretical Physics (KITP).

This work has made use of data from the European Space Agency (ESA) mission Gaia (\url{https://www.cosmos.esa.int/gaia}), processed by the Gaia Data Processing and Analysis Consortium (DPAC; \url{https://www.cosmos.esa.int/web/gaia/dpac/consortium}). Funding for the DPAC has been provided by national institutions, in particular the institutions participating in the Gaia Multilateral Agreement. This work is partly based on observations collected at the European Southern Observatory under ESO programme 114.27T5.001.
\end{acknowledgments}

\begin{contribution}
%%This section gives authors the space to recognize author contributions. The text inside this environment is NOT counted towards the total word quanta. At a minimum, manuscripts are expected to include this text:
%All authors contributed equally.

N.H. and S.S. jointly conceived the study, conducted the investigation, and wrote the manuscript, contributing equally to the work. O.I. led the analysis of the open-cluster MS$+$WD sample \citep{Ironi_2025}, while P.R. led the analysis of the Ba-enriched sample \citep{Rekhi_2024, Rekhi_2026}, including the comparison with GALAH presented in Appendix~\ref{app:BaAbundnace}. S.B.A. and H.W.R. contributed to the development of the underlying Ba enrichment survey presented in \citet{Rekhi_2024, Rekhi_2026}. All authors participated in scientific discussions and provided feedback on the manuscript.

%% But authors are expected to provide more specific details, e.g. 
%%
%%SC was responsible for writing and submitting the manuscript.
%%WWM came up with the initial research concept and edited the manuscript.
%%OTS obtained the funding and edited the manuscript.
%%EBF provided the formal analysis and validation. He also edited the manuscript.
%%GEH Supervised the undergraduates, wrote the software and administers the project github and Zenodo repositories.
%%
%% Authors can use the Contributor Role Taxonomy (CRediT) at
%% for ideas on how write a good statement tailored to their needs.

\end{contribution}

\facilities{Gaia}

\software{astropy \citep{Astropy_2013, Astropy_2018}, matplotlib \citep{Hunter_2007}, numpy \citep{Numpy_2006, Numpy_2011}, pysme \citep{Wehrhahn_2023}, scipy \citep{Virtanen_2020}, topcat \citep{Taylor_2005}}


\newpage

\appendix
\section{The Individual Components of IK~Peg}
\label{app:IKPeg}
\subsection{IK~Peg\,A}
\label{sec:IKPegA}

IK~Peg\,A is a $1.65\pm0.05$\,\msun\ A-type MS star, with an effective temperature of about 7700\,K, $\log g \approx 4.25$, and an age of 50--600\,Myr \citep{Wonnacott_1994}.
\citet{Cowley_1969} initially classified it as a marginal metallic-line (Am) star (`A8m:'), while \citet{Abt_1969} subsequently described it as ``definitely Am''. This classification refers to chemically peculiar A-type stars that exhibit deficiencies in calcium (Ca) and/or scandium (Sc), coupled with enhanced apparent abundances of iron-group and heavy elements \citep{Conti_1970}. These anomalous abundances are attributed to radiative levitation \citep{Michaud_1970, Watson_1970, Watson_1971, Smith_1971}. In this process, ions with numerous absorption lines (e.g., iron-group elements) absorb photons more efficiently and are pushed to the surface by radiation pressure, whereas lighter elements with fewer spectral lines settle below the atmosphere due to gravity. This mechanism is effective only if the star rotates slowly enough to maintain atmospheric stability. Such slow rotation is typically the consequence of tidal synchronization with a close companion in a binary system.

IK~Peg\,A also exhibits $\delta$~Scuti pulsations \citep{Kurtz_1978}, a phenomenon typical of A--F-type MS stars within the instability strip. These pulsations are driven by the helium partial ionization zone. However, due to the slow rotation of Am stars, gravitational settling is expected to deplete helium from this layer, suppressing the driving mechanism. At the projected rotational velocity of IK~Peg\,A ($v \sin i\approx 28.4$\,\kms, measured by Gaia), only $\sim 20\%$ of A--F-type stars in the instability strip show $\delta$~Scuti pulsations \citep{Gootkin_2024}. The fact that these pulsations are not suppressed also argues against a radiative levitation origin for the metallic abundance anomalies observed in IK~Peg\,A \citep{Wonnacott_1994}.

The observed abundances of IK~Peg\,A are inconsistent with those of typical Am stars. In a detailed analysis, \citet{Smalley_1996} showed that Ca and Sc are consistent with Solar values, while the iron-group elements follow the measured metallicity (see Figure~\ref{fig:IKPegAbundances}). In contrast, rare-Earth elements, which are usually strongly enhanced in Am stars, show no such enrichment. This pattern challenges an internal origin for the anomalous material. An alternative is that the enrichment is extrinsic, caused by mass transfer from the WD progenitor \citep{Landsman_1993, Smalley_1996}, analogous to the formation pathway of Ba stars.

The chemical abundance profile of IK~Peg\,A presents a notable discrepancy with standard \textit{s}-process pollution scenarios. While Ba, which is a representative second-peak element, is significantly enhanced, the first-peak elements exhibit only marginal (Sr) to negligible (Y, Zr) enrichment (see Figure~\ref{fig:IKPegAbundances}). Such a high ratio of heavy-to-light \textit{s}-process elements ([hs/ls]) is indicative of high neutron exposure. We note that the reported marginal Sr enhancement, derived from a limited number of spectral lines \citep{Smalley_1996}, is statistically consistent with no enhancement within $2\sigma$ and may reflect measurement uncertainty. Crucially, high [hs/ls] ratios are typically characteristic of low-metallicity environments \citep[e.g.,][]{Bisterzo_2011}, a condition not met by the solar-metallicity IK~Peg system. A modern re-analysis of the IK~Peg\,A spectrum might resolve this inconsistency.

\subsection{IK~Peg\,B}
\label{sec:IKPegB}

The system's high mass function, $f(M)\equiv(M_2 \sin i)^3/(M_1+M_2)^2=0.161$\,\msun\ \citep{Batten_1989}\footnote{More recent estimates place the mass function at $f(M)=0.219\pm0.004$ \citep{Vennes_1998}.}, combined with the non-detection of the companion in the optical spectrum, led \citet{Trimble_1969} to suggest that it may be a compact object.
Assuming a primary mass of $1.7$\,\msun\ implies a companion mass of $>1.1$\,\msun\ \citep{Landsman_1993}.
The detection of extreme ultraviolet (EUV) and X-ray flux originating from the system suggested the companion is a WD. This hypothesis was subsequently confirmed by a direct far-UV (FUV) spectrum of the WD, revealing a massive, hydrogen-dominated (DA) WD  \citep{Wonnacott_1993}. Based on an atmospheric fit to low-resolution FUV spectra obtained with the International Ultraviolet Explorer (IUE), \citet{Landsman_1993} estimated an effective temperature of $35{,}500\pm500$\,K and $\log g=9.0^{+0.15}_{-0.3}$, implying a mass of $1.15^{+0.05}_{-0.15}$\,\msun\ ($T_\text{eff}=35{,}350\pm150$\,K, $\log g=8.95\pm0.1$, and $1.185\pm0.055$\,\msun\ according to \citet{Vennes_1998} using Hipparcos parallax). The orbital separation is then $3.1 \times 10^7$\,km, or about 0.21\,au.

The WD in the IK~Peg system has a cooling age of $\sim 10^8$\,yr. Assuming a massive progenitor ($\approx 8$\,\msun) with an MS lifetime of $5 \times 10^7$\,yr, the system's total age implies its origin cluster should be identifiable. However, no such cluster has been found. This absence led \citet{Landsman_1993} to suggest that IK~Peg originated as a triple system, where the present-day massive WD was formed by the merger of an inner binary.

\section{Ba Abundance Estimate Uncertainty}
\label{app:BaAbundnace}

\begin{figure}
    \centering
    \includegraphics[width=\columnwidth]{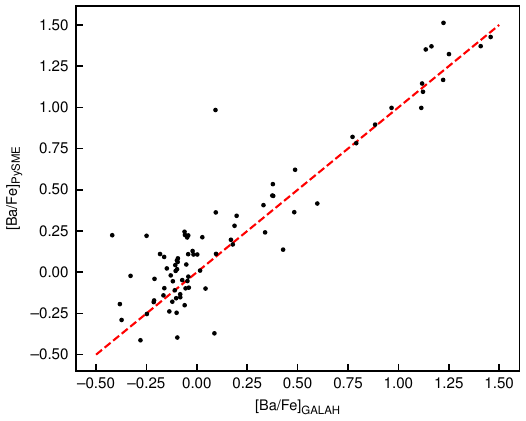}
    \caption{Comparison of the [Ba/Fe] abundances for the \citet{Rekhi_2024} sample as reported in GALAH DR3 versus those derived using \textsc{pysme} (following the methodology of \citealt{Rekhi_2026}). The red dashed line indicates the 1:1 relation for reference.}
    \label{fig:BaFe_PySME_vs_GALAH}
\end{figure}

To quantify the typical uncertainties in the Ba abundance estimates, we compared the values reported in GALAH DR3 \citep{Buder_2021} for the \citet{Rekhi_2024} sample with those derived using \textsc{pysme}, following the methodology of \citet{Rekhi_2026}. The comparison includes 80 systems, and is shown in Figure~\ref{fig:BaFe_PySME_vs_GALAH}. For each target, all stellar parameters were fixed to the GALAH DR3 values, and only the Ba abundance was varied. We adopted the solar iron abundance used by GALAH DR3, $A(\mathrm{Fe})_\odot = 7.38$ \citep{Buder_2021}. 

\begin{figure}
    \centering
    \includegraphics[width=\columnwidth]{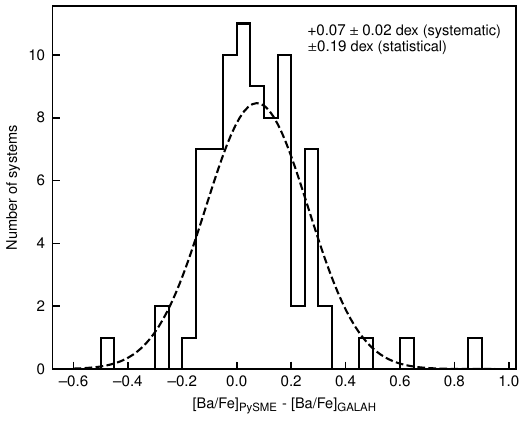}
    \caption{Distribution of the differences in [Ba/Fe] abundances for the \citet{Rekhi_2024} sample, comparing values reported in GALAH DR3 with those derived using \textsc{pysme} (following the methodology of \citealt{Rekhi_2026}). The distribution reveals a systematic offset of $+0.07\pm0.02$\,dex and a scatter of $0.19$\,dex.}
    \label{fig:BaFe_PySME_vs_GALAH_diff}
\end{figure}

The resulting distribution of [Ba/Fe] differences (Figure~\ref{fig:BaFe_PySME_vs_GALAH_diff}) exhibits a scatter of $0.19$\,dex and a systematic offset of $+0.07\pm0.02$\,dex. Given the sample size, the uncertainty on the mean offset is reduced relative to the dispersion of individual measurements, such that the offset is determined more precisely than the intrinsic scatter, with a $3\sigma$ upper bound $\lesssim 0.15$\,dex. Consequently, the uncertainty associated with correcting for this offset is small compared to the typical measurement uncertainties of individual systems.

\section{IK~Peg--type Systems}
\label{app:IKPegType}

\begin{figure}
    \centering
    \includegraphics[width=\columnwidth]{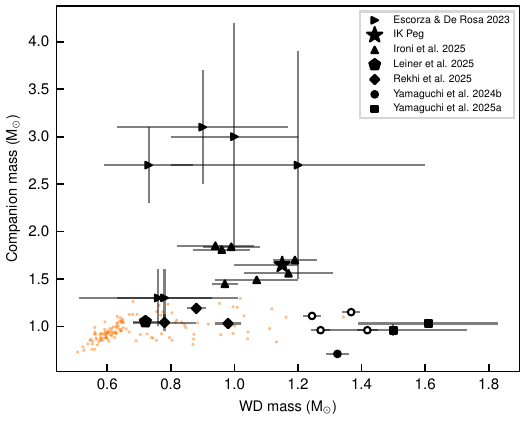}
    \caption{Component masses of the candidate IK~Peg--type systems listed in Table~\ref{tab:ikpeglike}. Open markers indicate lower limits of the WD masses. For reference, the full (albeit biased) sample of \citet{Rekhi_2024, Rekhi_2026} is plotted as orange points.}
    \label{fig:IKPegType_Masses}
\end{figure}

\begin{figure}
    \centering
    \includegraphics[width=\columnwidth]{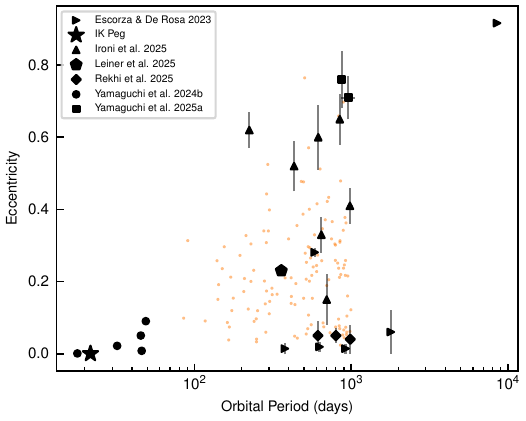}
    \caption{Period--eccentricity plot of the candidate IK~Peg--type systems listed in Table~\ref{tab:ikpeglike}. For reference, the full (albeit biased) sample of \citet{Rekhi_2024, Rekhi_2026} is plotted as orange points.}
    \label{fig:IKPegType_Pe}
\end{figure}

Table~\ref{tab:ikpeglike} lists potential IK~Peg--type systems identified in this work. The component masses and orbital properties of these candidates are illustrated in Figures~\ref{fig:IKPegType_Masses} and \ref{fig:IKPegType_Pe}, superimposed on the full sample from \citet{Rekhi_2024, Rekhi_2026} for comparison. We emphasize that due to the inherent selection biases in both the candidate list and the parent sample, robust statistical inferences regarding the differences between these populations cannot currently be drawn.

\startlongtable
\begin{deluxetable*}{lrrrrrrc}
\tablecaption{Candidate IK~Peg--type systems \label{tab:ikpeglike}}
\tablehead{
\colhead{Name} & \colhead{Gaia DR3 ID} & \colhead{$M_\text{MS}$ (\msun)} & \colhead{$M_\text{WD}$ (\msun)} & \colhead{Orbital period (d)} & \colhead{Eccentricity} & \colhead{[Fe/H]} & \colhead{Note}
}
\startdata
IK~Peg & 1787127787263968640 & $1.65\pm0.05$ & $1.15_{-0.15}^{+0.05}$ & $21.72168\pm0.00009$ & $0$ & $+0.17\pm0.17$ & 1,2,4 \\
\hline
J2117+0332 & 2692960678029100800 & $1.11\pm0.03$ & $\geq 1.244\pm0.027$ & $17.9239\pm0.0001$ & $0.0007\pm0.0002$ & $-0.09\pm0.05$ & 1,5\\
J1111+5515 & 843829411442724864 & $1.15\pm0.02$ & $\geq 1.367\pm0.029$ & $32.1494\pm0.0022$ & $0.0217\pm0.0003$ & $-0.09\pm0.06$ & 1,5\\
J1314+3818 & 1522897482203494784 & $0.71\pm0.01$ & $1.324\pm0.037$ & $45.5150\pm0.0047$ & $0.0503\pm0.0003$ & $-0.33\pm0.04$ & 1,5\\
J2034$-$5037 & 6475655404885617920 & $0.96\pm0.02$ & $\geq 1.418\pm0.033$ & $46.1147\pm0.0006$ & $0.0079\pm0.0002$ & $-0.16\pm0.06$ & 1,5\\
J0107$-$2827 & 5033197892724532736 & $0.96\pm0.03$ & $\geq 1.271\pm0.031$ & $49.0063\pm0.0008$ & $0.0901\pm0.0005$ & $+0.08\pm0.07$ & 1,5\\
\hline
--- & 6091100873778091392 & $1.04\pm0.09$ & $0.78\pm0.10$ & $987.29\pm65.09$ & $0.04\pm0.04$ & $-0.11\pm0.08$ & 2,6 \\
--- & 5722712938857848192 & $1.19\pm0.06$ & $0.88\pm0.03$ & $614.79\pm3.52$ & $0.05\pm0.04$ & $0.19\pm0.05$ & 2,6 \\
--- & 5550946678313327744 & $1.03\pm0.06$ & $0.98\pm0.04$ & $800.09\pm22.16$ & $0.05\pm0.02$ & $0.06\pm0.07$ & 2,6 \\
\hline
--- & 1386979565629462912 & $1.03\pm0.05$ & $1.61\pm0.22$ & $873.1\pm9.1$ & $0.76\pm0.08$ & ${\sim}\,-1.4$ & 2,7\\
--- & 5648541293198448512 & $0.96\pm0.06$ & $1.50\pm0.23$ & $960\pm100$ & $0.71\pm0.06$ & $0.04\pm0.05$ & 2,7\\
\hline
HD 123949 & 6290109626537096192 & $1.30\pm0.30$ & $0.78\pm0.15$ & $8544.00\pm12.00$ & $0.92\pm0.00$ & $-0.23$ & 2,8 \\
HD 49641 & 3127391779597616000 & $2.70\pm1.20$ & $1.20\pm0.40$ & $1793.00\pm21.00$ & $0.06\pm0.06$ & $-0.30$ & 2,8 \\
HD 31487 & 260022272200061824 & $2.50\pm0.20$ & $1.59\pm0.22$ & $1063.80\pm0.40$ & $0.04\pm0.02$ & $-0.04$ & 2,8 \\
HD 92626 & 5364553966782362752 & $3.10\pm0.60$ & $0.90\pm0.27$ & $921.70\pm1.70$ & $0.01\pm0.01$ & $-0.15$ & 2,8 \\
HD 44896 & 2891763860781890176 & $3.00\pm1.20$ & $1.00\pm0.20$ & $629.00\pm1.10$ & $0.02\pm0.01$ & $-0.25$ & 2,8 \\
HD 199939 & 2162250505790243840 & $2.70\pm0.40$ & $0.73\pm0.14$ & $585.39\pm0.09$ & $0.28\pm0.01$ & $-0.22$ & 2,8 \\
HD 24035 & 4641654174211281792 & $1.30\pm0.30$ & $0.76\pm0.25$ & $378.00\pm0.50$ & $0.01\pm0.02$ & $-0.23$ & 2,8 \\
\hline
Candidate \#6 & 183325907326150528 & $1.85\pm0.03$ & $0.94\pm0.12$ & $617.54\pm12.50$ & $0.60\pm0.09$ & $0.05\pm0.03$ & 3,9 \\
Candidate \#110 & 2166849556774987520 & $1.84\pm0.02$ & $0.99\pm0.09$ & $224.28\pm1.36$ & $0.62\pm0.05$ & $0.15\pm0.02$ & 3,9 \\
Candidate \#154 & 3451094801741563264 & $1.49\pm0.02$ & $1.07\pm0.13$ & $983.40\pm42.60$ & $0.41\pm0.05$ & $0.14\pm0.02$ & 3,9 \\
Candidate \#183 & 5242761476090953728 & $1.45\pm0.02$ & $0.97\pm0.04$ & $699.25\pm7.79$ & $0.15\pm0.07$ & $0.06\pm0.06$ & 3,9 \\
Candidate \#260 & 5870278497578560128 & $1.70\pm0.03$ & $1.19\pm0.07$ & $646.42\pm3.98$ & $0.33\pm0.05$ & $0.03\pm0.04$ & 3,9 \\
Candidate \#262 & 5930051213822024704 & $1.56\pm0.03$ & $1.17\pm0.14$ & $847.39\pm20.73$ & $0.65\pm0.07$ & $0.11\pm0.04$ & 3,9 \\
Candidate \#277 & 6071002793352423424 & $1.81\pm0.02$ & $0.96\pm0.09$ & $433.09\pm3.42$ & $0.52\pm0.07$ & $0.18\pm0.04$ & 3,9 \\
\hline
WOCS~14020 & 604906531159503616 & $\sim 1.05$ & $0.72_{-0.04}^{+0.05}$ & $358.9$ & $0.23$ & $\sim 0$ & 3,10 \\
\enddata
\tablecomments{(1) Massive WD in $\sim 0.1$\,au orbit.
                (2) Ba-enhanced.
                (3) Inconsistent with cluster's age.
                (4) $M_\text{MS}$ from \citealt{Wonnacott_1994}; $M_\text{WD}$ from \citealt{Landsman_1993}; orbital period and eccentricity from \citealt{Harper_1927, Harper_1935, Vennes_1998}; metallicity from \citealt{Smalley_1996}.
                (5) \citealt{Yamaguchi_2024_WidePCEBs}.
                (6) \citealt{Rekhi_2026}.
                (7) \citealt{Shahaf_2024, Yamaguchi_2025_Barium}.
                (8) Giant Ba stars from \citealt{Escorza_2023}.
                (9) \citealt{Ironi_2025}.
                (10) \citealt{Leiner_2025}.}
\end{deluxetable*}

%% For this sample we use BibTeX plus aasjournalv7.bst to generate the

\bibliography{ikpeg}{}
\bibliographystyle{aasjournalv7}



\end{document}